\documentstyle[aps,prl,epsf,multicol]{revtex}
\begin{document}
\draft

\title{{Reply to comment on ``Dynamic scaling in the spatial
distribution of persistent sites''}}

\author{ G.\ Manoj and P.\ Ray}

\address{ The Institute of Mathematical Sciences, C. I. T. Campus,
Taramani, Madras 600 113, India}

\maketitle
	
\begin{multicols}{2}	
In their comment\cite{ben} on our paper \cite{manoj}, Ben-Naim and 
Krapivsky  
claims to have presented numerical evidence against our conclusions.
We show that their claims are not valid for the following
reasons, which we discuss here. 
\newline (i)  The arguments given in \cite{ben} against 
non-universality are incorrect.
\newline (ii) The apparent disagreement between the results
arises possibly from a difference in the
initial conditions used.
\newline (iii) We present new numerical results, which support
our earlier conclusions.

In \cite{ben}, the authors emphasise that their scaling function
is independent of all initial conditions. This
is a surprising result, because  
unlike the case of particle density $n(t)$, the prefactor $\Gamma$
appearing in the asymptotic expression $P(t)=\Gamma t^{-\theta}$
depends on the initial density $n_{0}$. 
To show this, we first note that for an initial particle density $n_{0}$, 
the effects of the annihilation 
reaction will be felt only after a time interval $t_{0}\sim (Dn_{0}^{2})^{-1}$.
For $t\gg t_{0}$, $n(t)\sim t^{-1/2}$ and so
$P(t)=P(t_{0})(t/t_{0})^{-\theta}$.
On the other hand, for $t\ll t_{0}$, $n(t)\approx n_{0}$ so that
$P(t)\simeq P(0)e^{-\alpha n_{0}\sqrt{Dt}}$
in this regime (following an argument given in \cite{redner}). 
Here $P(0)$ is the initial density of persistent sites
and $\alpha$ is a numerical constant. 
Combining the two,
one finds $\Gamma\sim P(0){n_{0}}^{-2\theta}$, which we have 
verified in simulations. (Similar arguments 
can be used to show that the prefactor for particle
density $n(t)$ is independent of $n_{0}$).

It is easy to see that $1/2$ cannot be a universal value for
the dynamical exponent $z$ proposed in \cite{manoj}, contrary to the 
claims in \cite{ben}. To show this, let us consider the 
$q$-state Potts model in the limit $q\to\infty$, the dynamics
of which is given by $A+A\to A$ reaction which also has
$n(t)\sim t^{-1/2}$. It is known from
the exact solution \cite{derrida} that $\theta=1$ for this model.
If we now follow the arguments in \cite{ben}, we find that
their scaling function ${\cal F}(x)\sim$ constant for $x\ll 1$
and decays exponentially for $x\gg 1$ (where $x=lt^{-1/2}$).
Recomputing the persistent fraction $P(t)$ using this distribution
yields the inconsistent result $P(t)\sim t^{-1/2}$.

Coming to the numerics, first of all, we would like to point out that 
$L(t)$ curves in \cite{ben} for $n_{0}=0.5$ and $n_{0}=0.8$ are
almost identical, unlike the corresponding plots given 
in our paper. This is possibly due to a 
difference in the initial condition: the entire lattice
being taken as persistent at $t=0$ i.e, $P(0)=1$ 
instead of $P(0)=1-n_{0}$ as in \cite{manoj}.
In the former case, the fragmentation of the lattice into
clusters of persistent and non-persistent sites 
is delayed by one MC step (for $D=1/2$).  
Consequently, the effective particle density is only 
${n_{0}}^{*}=n_{0}-\delta n_{0}$,
where $\delta n_{0}$ is the decrease in density over one time step.
We have found numerically that for $n_{0}=0.2$, 0.5 and 0.8, 
${n_{0}}^{*}\simeq 0.15$, 0.25 and 0.29 respectively. In
this small range, it is difficult to make any inference regarding 
the universality of exponents. 

\begin{figure}[b]
\narrowtext
\epsfxsize=2.2in
\epsfbox{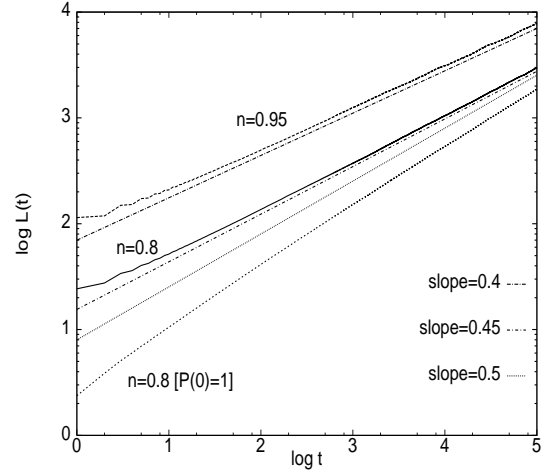}
\vspace{-1.0cm}
\caption{
$L(t)$ is plotted against t 
for $n_{0}=0.95$ and $n_{0}=0.8$. For 
comparison, the plot for $n_{0}=0.8$ with the initial 
condition $P(0)=1$ is also shown, which 
is identical to the $n_{0}=0.8$ plot 
in the comment. Our system size is $10^{5}$
and the results were averaged over 100 different initial
realizations. We have also done one $10^{6}$ size lattice,
with identical results. Note that for large $n_{0}$, the
asymptotic behaviour sets in very early.
} 
\end{figure}

Finally, we present more numerical results (Fig. 1)
which endorses the conclusions in our paper. 
In particular,
we see that as $n_{0}\to 1$,  
$z\simeq \frac{\theta}{n_{0}}$ (Table I), in
agreement with our earlier observations.

\begin{table}
\narrowtext	
\begin{tabular}{cccc}
$n_{0}$ & $\theta$ & $z$  & $\frac{\theta}{n_{0}}$\\
\hline
0.8 & 0.37515(1) & 0.44916(2) & 0.46894\\
0.95 & 0.37573(1) & 0.39517(2) & 0.39551\\
\end{tabular}
\caption{Exponents measured
from simulations on $10^{5}$ size systems. 
Figures in brackets represent statistical error
in the last decimal place.}
\end{table}

\end{multicols}

\end{document}